# Digital Image Watermarking for Arbitrarily Shaped Objects Based On SA-DWT

A.Essaouabi[1], E.Ibnelhaj[2], F.regragui[1]

[1] Department of physics, LIMIARF Laboratory, Mohammed V University
Rabat, Avenue Ibn Battouta B.P.1014 RP, Morocco

[2] Institut National of Posts and Telecommunications,
Av. Allal Al Fassi-Madinat Al Irfane Rabat,Morocco

## Abstract

Many image watermarking schemes have been proposed in recent years, but they usually involve embedding a watermark to the entire image without considering only a particular object in the image, which the image owner may be interested in. This paper proposes a watermarking scheme that can embed a watermark to an arbitrarily shaped object in an image. Before embedding, the image owner specifies an object of arbitrary shape that is of a concern to him. Then the object is transformed into the wavelet domain using in place lifting shape adaptive DWT(SADWT) and a watermark is embedded by modifying the wavelet coefficients. In order to make the watermark robust and transparent, the watermark is embedded in the average of wavelet blocks using the visual model based on the human visual system. Wavelet coefficients $n$ least significant bits (LSBs) are adjusted in concert with the average. Simulation results shows that the proposed watermarking scheme is perceptually invisible and robust against many attacks such as lossy compression (e.g. JPEG, JPEG2000), scaling, adding noise, filtering, etc.

## 1. Introduction

With the rapid growth of Internet technologies and wide availability of multimedia computing facilities, the enforcement of multimedia copyright protection becomes an important issue. Digital watermarking, one of the popular approaches considered as a tool to achieve this goal, is a technique based on embedding a specific mark or signature into the digital products. Many watermark algorithms have been proposed to address this issue of copyright protection. Cox et al. [1] propose a DCT based spread spectrum watermarking technique. A pseudo-random sequence is embedded into the significant DCT coefficients and is retrieved by calculating the similarity function of the original watermark and extracted watermark. Su et al. [2] proposes a wavelet-based watermark algorithm. Based on the principle of multithreshold wavelet codec (MTWC), the method searches the significant wavelet coefficients to embed the watermark in order to increase the robustness. The embedding strength in each subband is determined by the threshold of the subband. Polilchuk and Zeng [3] propose two kinds of adaptive watermarking methods. One is based on discrete cosine transform (IA-DCT), the other is based on discrete wavelet transform (IA-W). The watermark is embedded according to the JND threshold. Kaewkamnerd et al. [4] propose a wavelet based adaptive watermarking scheme. The human visual system (HVS) is employed to determine the weighting function T(x,y) to control the watermark casting process.

The problem with the current watermarking algorithms is that most of them embed a watermark in the entire image without taking the content of the image into account. In some situations, the image owner may be more interested in an object of an image than the whole image, so it's desirable to embed a watermark in the object to protect it better. To address this problem, Guo and Georganas [5] propose a watermarking scheme that can embed a watermark to an arbitrarily shaped object in an image. Before embedding, the image owner specifies an object of arbitrary shape that is of a concern to him. Then the object is transformed into the spectrum domain using shape adaptive DCT and a watermark is embedded by modifying the spectrum coefficients in an additive way. But the watermark can be damaged by a wavelet-based image codec. Therefore, this method limits their applications in the context of JPEG2000 due to the fact that the wavelet transform is playing an important role in JPEG2000. Kong Yu and Liu [6] propose a novel blind object watermarking scheme for images using SA-DWT.

To make the watermark robust and perceptual invisible, the watermark is embedded in the weighting mean of the wavelet blocks using the quantisation visual model based on HVS. The visual model takes into account the brightness sensitivity and texture sensitivity. Watermark detection is accomplished without the original, unwatermarked object by using statistical detection technique. This technique is robust against many attacks such as lossy image/video compression (e.g. JPEG, JPEG2000), scaling, adding noise, filtering, D/A and A/D conversion, etc.

In this way, we propose in this paper a new blind watermarking scheme of images based on the in place lifting SA-DWT. The watermark signal is embedded in the wavelet coefficients $n$ LSBs. Unlike most watermark schemes, watermark embedding is performed by modulating the average of the wavelet coefficients instead of the individual coefficients in the wavelet block. Visual model is employed to achieve the best tradeoff between transparent and robustness to signal processing. Watermark detection is accomplished without the original. Experimental results demonstrate that the proposed watermarking scheme is perceptually invisible and robust against unintentional and intentional attacks such as lossy image compression (e.g. JPEG, JPEG2000), scaling, adding noise, filtering.

The rest of the paper is organized as follows: Section 2 briefly introduces the in place lifting shape adaptive DWT. In section 3, the proposed scheme is introduced and section 4 presents some experimental results. Finally, this paper concludes with section 5.

## 2. In-Place Lifting SA-DWT

Given an arbitrarily shaped object with shape mask information, with in place lifting SA-DWT[7], the number of transformed coefficients is equal to the number of pixels in the arbitrarily shaped segment image, and the spatial correlation across subbands is well preserved. Fig. 1 illustrates the result of one-level wavelet decomposition of an arbitrarily shaped object.

The in-place lifting DWT implementation has special implications for the SA-DWT[8], which can best be understood visually as shown in Fig.1. As the SA-DWT is performed, the spatial domain shape mask remains intact with no requirement to derive a shape mask for each subband. How the subbands are arranged in this pseudo-spatial domain arrangement is shown in Fig. 2(a). Each subband can in fact be extracted from the interleaved subband arrangement using the lazy wavelet transform (LWT) [9]. After the one-level SA-DWT is performed, the LL1 subband can be extracted using a coordinate mapping from the interleaved subband coordinates (i,j) to the LL1 subband coordinates ($i_{LL1}, j_{LL1}$) as follows:

$$(i_{LL1}, j_{LL1}) \leftarrow ([i/2], [j/2]) \qquad (1)$$

Similarly, the mapping for the HL1 subband is
($i_{HL1}, j_{HL1}$) $\leftarrow$ ([i/2]+1, [j/2]); for the LH1 subband ($i_{LH1}, j_{LH1}$) $\leftarrow$ ([i/2], [j/2]+ 1); and for the HH1 subband ($i_{HH1}, j_{HH1}$) $\leftarrow$ ([i/2] + 1, [j/2] + 1).

After the first level of the SA-DWT, the interleaved subband arrangement is made up of 2 × 2 basic blocks of coefficients. As shown in the left side of Fig. 2 (b), the top-left coefficient of each block is an LL1 subband coefficient, the top-right coefficient is an HL1 subband coefficient, and so on the second level SA-DWT is performed by first extracting the LL1 subband using the coordinate mapping (1) and then performing the one-level SA-DWT using the LL1 subband as the new input. The output is the four interleaved subbands, LL2, HL2, LH2, and HH2. This is then placed back into the original interleaved subband arrangement where the LL1 coefficients were extracted from. This creates a two-level interleaved subband arrangement. As shown in the middle of Fig.2(b), the two-level interleaved subband arrangement is made of a basic 4×4 coefficient block, with the top-left coefficient of each block being an LL2 coefficient. The coordinate mappings to extract the second and subsequent level subbands are simply derived by applying the one level coordinate mappings iteratively to the LL subband coordinate mapping from the previous level.

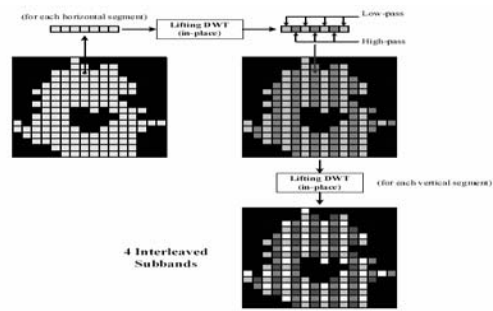

Fig. 1  One-level, two-dimensional SA-DWT using in-place lifting DWT implementation

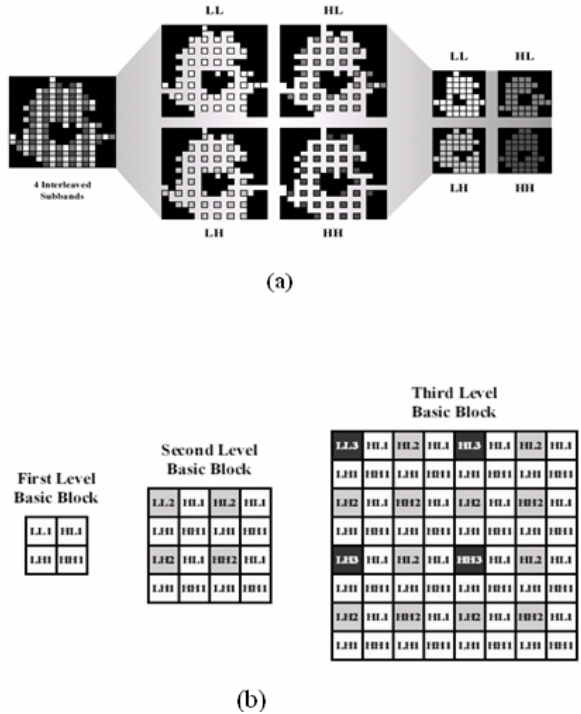

(a)

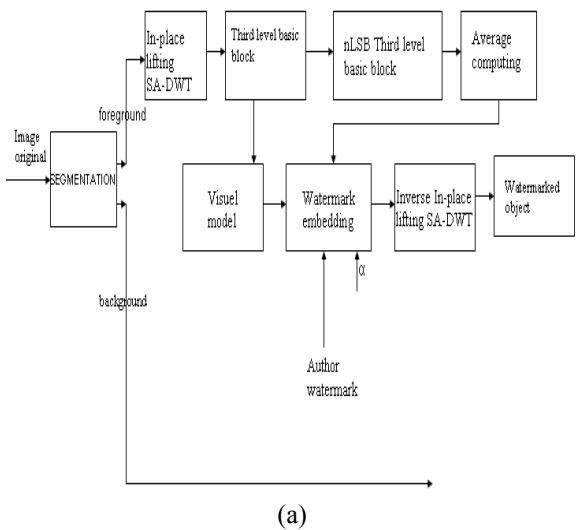

(b)

Fig. 2  (a) Interleaved subband abstraction  (b) Basic group of coefficients for each level of in-place DWT.

# 3. Proposed Watermarking Scheme

A content-based watermarking system for content integrity protection is illustrated in Fig  3.

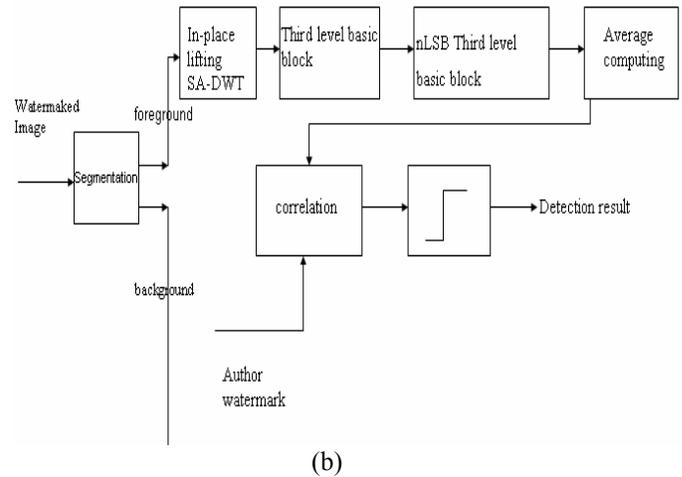

(b)

Fig. 3  Block diagrams for the proposed watermarking scheme. (a) Watermark embedding. (b) Watermark detection.

## 3.1 Watermark Embedding

Fig.3(a) shows the watermarking embedding procedure. First instead, the original image is segmented into foreground (object) and background and we apply the three levels in   place lifting SA-DWT to foreground object .Then we  take each third level basic block (see Fig 2(b)). NxN is the size of the matrix wavelet block and $I_i(k)$ is the ith wavelet coefficient in the kth wavelet block where  i $\in$ [1, N×N].

The rest of the watermarking embedding procedure is presented in the following. The    n LSBs  of  $I_i(k)$ is defined as :

$$\hat{I}_i(k) = \mathrm{mod}(I_i(k), 2^n) \qquad (2)$$

The average of the wavelet block is defined as follows:

$$Average \quad (k) = \frac{\sum_{i=1}^{N \times N} \hat{I}_i(k)}{N \times N} \qquad (3)$$

In the proposed watermarking, we choose the blocks with an average value different from zero.

If a few of   $I_i(k)$ are changed by $\Omega$ due to some distortions, the average of the wavelet block will only have a small change[6]. Assuming that $I'_i(k)$ is the ith wavelet coefficient in the kth wavelet block after the watermark embedding, $\hat{I}'_i(k)$ is the n LSBs of $I'_i(k)$ and Average'(k ) is the average of $\hat{I}_i'(k)$ in the kth wavelet block accordingly. The watermark $W$, consisting of a binary pseudo random sequence, $W(k) \in \{-1, 1\}$, is embedded by adjusting the average of wavelet blocks in this way :

$$Average'(k) \in \begin{cases} [0, 2^{n-1}), if & W(k) = -1 \\ [2^{n-1}, 2^n), if & W(k) = 1 \end{cases} \quad (4)$$

To adapt the watermark sequence to the local properties of the wavelet block, we use the model based on HVS in the watermark system. The visual model function $Vm(k)$ is defined as:

$$Vm(k) = brightness(k) \times texture(k)^\beta \quad (5)$$

Where

$$texture(k) = \frac{\sum_{i=1}^{N \times N} [brightness(k) - I_i(k)]^2}{N \times N}$$

$$brightness(k) = \frac{\sum_{i=1}^{N \times N} I_i(k)}{N \times N}$$

$\beta$ is a parameter used to control the degree of texture sensitivity. This visual model function indicates that the human eye is less sensitive to noise in the highly bright and the highly textured areas of the image. Hence, the wavelet blocks are divided into two parts depending on the value of $Vm(k)$: high activity wavelet block and low activity wavelet block. For simplicity, the threshold $Tc$ is set to the average of $Vm(k)$. The following function can be applied to distinguish high or low activity wavelet block:

$$T(k) = sign(Vm(k) - Tc) \quad (6)$$

Considering the tradeoff between robustness and transparency, the proposed watermark embedding algorithm can be formulated as follows:

$$I_i'(k) = I_i(k) + \alpha W(k) F_i(k) [2^{n-2-S(k)} + T(k) \times 2^{n-3}] \quad (7)$$

where $\alpha$ is a scaling factor used to control the strength of the inserted watermark. The flag function is defined as follows:

$$F_i(k) = sign((2^{n-1} - \hat{I}_i(k)) \times W(k)) \quad (8)$$

Where

$$sign(x) = \begin{cases} 1 & if \quad x \geq 0 \\ -1 & if \quad x < 0 \end{cases}$$

The strength function is defined as follows:

$$S(k) = sign(X(k)) \quad (9)$$

Where

$$X(k) = (2^{n-1} - Average(k)) \times W(k)$$

Details concerning the flag function and the strength function are described in table 1.

*Table 1.* The detailed results of $F_i(k)$ and $S(k)$

| W(k) | $2^{n-1} - \hat{I}_i(k)$ | $2^{n-1} - Average(k)$ | F(k) | S(k) |
|------|------|------|------|------|
| -1 | >0 | >0 | -1 | -1 |
| -1 | ≤0 | ≤0 | -1 | -1 |
| 1 | >0 | >0 | 1 | 1 |
| 1 | ≤0 | ≤0 | 1 | 1 |

In light of the above, the $n$ LSBs of wavelet coefficients have been adjusted by using equation (6). Naturally, their average has been updated depending on the requirement of $W(k)$ as show in equation (3). In other word, the watermark has been embedded.

## 3.2 Watermark Extraction and Detection

The watermark sequence can be extracted without the original object. From the process of the watermark embedding, we can obtain the watermarked objects by applying the function of equation (3). Thus, for a given watermarked object, the watermark can be extracted as

$$W'(k) = \begin{cases} -1, & if \quad Average(k) \in [0, 2^{n-1}) \\ 1, & if \quad Average(k) \in [2^{n-1}, 2^n) \end{cases} \quad (10)$$

In order to detect the watermark $W'$ extracted from the watermarked object, First instead we evaluate the detector response (or similarity of $W'$ and $W$) as :

$$\rho(W',W) = \frac{\sum_{k=1}^{L} W'(k) \times W(k)}{\sum_{k=1}^{L} \|W'(k)\|^2} = \frac{\sum_{k=1}^{L} W'(k) \times W(k)}{L} \qquad (11)$$

Where, $L$ is the length of the watermark signal. The Threshold $T\rho$ is set so as to minimize the sum $p$ of the probability of error detection and the probability of false alarm. if $\rho \geq T\rho$, we considered the watermark is present, otherwise absent.

## 4. Experiments Results

Simulations are carried out for several standard monochrome images as shown in Fig .4 but only report result in detail for 704 x 480 akiyo. In our experiments, the parameters considered are: the threshold $T\rho = 0.1$, $\beta = 0.318$, $n = 5$, $N=8$ wavelet-*level* = 3, wavelettype ='haar', L =1700 and scaling factor $\alpha$ =0.3.

In order to test the performance of the proposed watermarking scheme, 200 watermarks were randomly generated.

The PSNR result between the original object and the watermarked object is 39.26 dB. As shown in Fig. 5, the watermark is perceptual invisible and the object with watermark appears visually identical to the object without watermark. In Fig. 6 the absolute difference between the original object and the watermarked one, it is evident that there is no watermark embedded in the region outside the object. Fig. 7 shows the response of the watermark detector to 200 randomly generated watermarks of which only one matches the watermark present. The response to the correct watermark (i.e. number 100) is much higher than the responses to incorrect watermarks.

To evaluate the robustness of our scheme against unintentional and intentional attacks, we test the watermarked object with JPEG, JPEG2000, scaling, adding noise, filtering, and multiple watermarking attack. Added experiment results for other images are listed in Table 2.

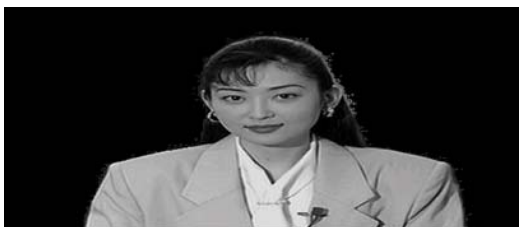

(a)

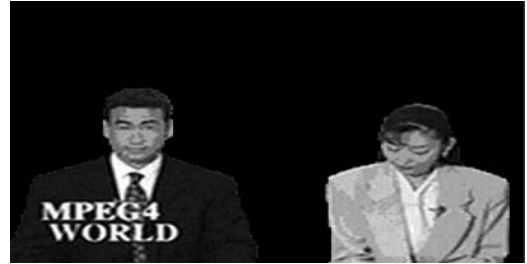

(b)

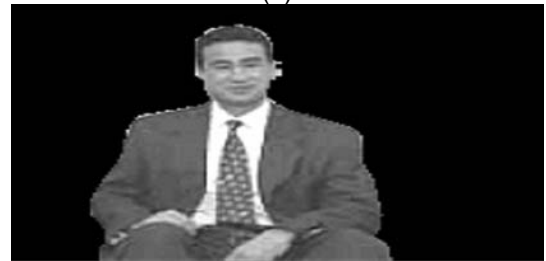

(c)

Fig. 4 (a)Original image (object) akiyo, (b)Original image (object) News, (c)Original image (object) Scene.

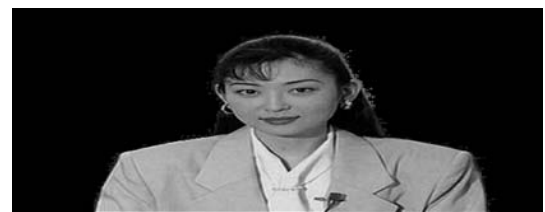

Fig. 5 Watermarked object akiyo (PSNR=39.26 dB)

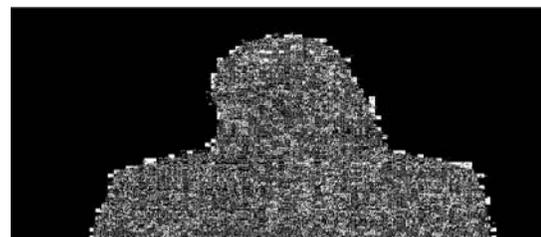

Fig. 6 Absolute difference between the original object and the watermarked

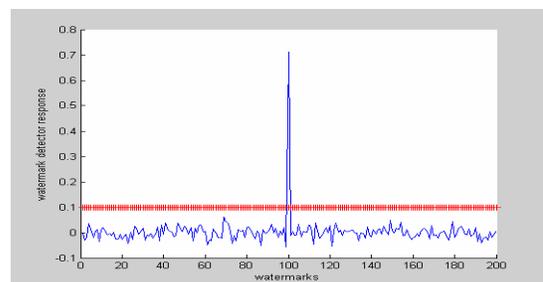

Fig. 7 Detector response of the watermarked object akiyo for 200 randomly generated watermark

### 4.1 JPEG Compression Distortion

JPEG is a widely used compression format and the watermark should be resistant to it. As shown in Fig.8, with the decreasing of the quality of the JPEG compressed object, the response of the watermark detector also decreases. We have found that the proposed watermark can survive even with quality factor of 40% (see Fig.9), although the object is visibly non distorted (see Fig.10)

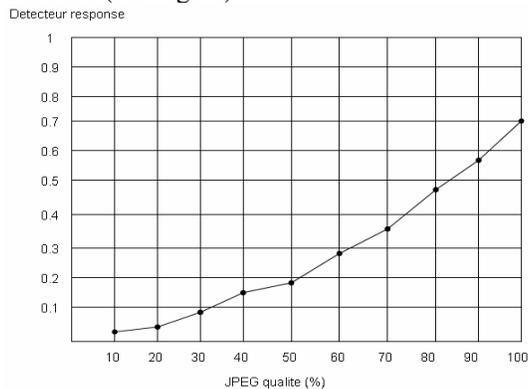

Fig. 8  Watermark detector response on the decreasing of the quality of the JPEG compressed object' akiyo'

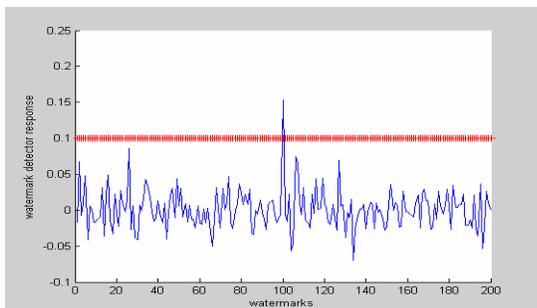

Fig. 9  Detector response to a JPEG compression copy of the watermarked object 'Akiyo' with 40% quality.

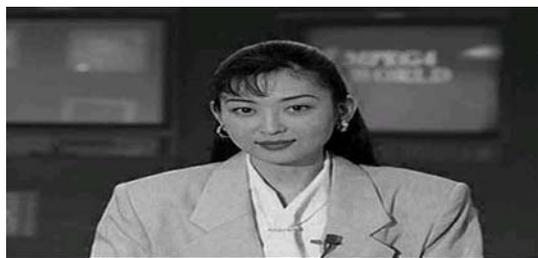

Fig. 10  JPEG compressed copy of the watermarked object 'Akiyo' with 40% quality.

### 4.2 JPEG2000 Compression Distortion

JPEG2000 is the new generation compression standard, which is based on wavelet transform. In our experiments, we test the watermarked object with JPEG2000 compression. The detector response of the watermarked object Akiyo after the JPEG2000 compression with 75% quality is 0.3729. Fig. 11 shows the object after the JPEG2000 compression with 65% quality, which results in very significant distortion. The response of the watermark detector in this case is 0.2053, which is still above the threshold Tc (see Fig.12).

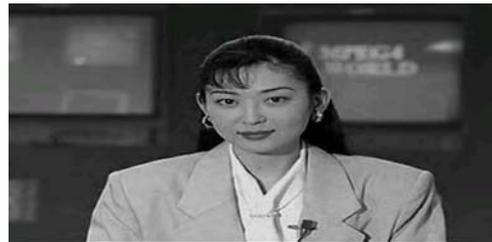

Fig. 11  JPEG2000 compression copy of the watermarked object 'Akiyo' with 65% quality.

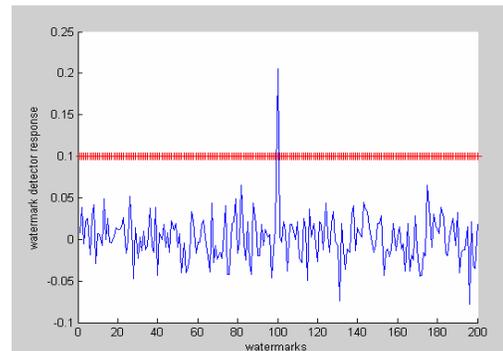

Fig. 12  Detector response to a JPEG2000 compression copy of the watermarked object 'Akiyo' with 65% quality.

### 4.3 Adding Noise

Noise is one of common distortions in image processing and transmission. In the experiment, we add 30% uniform noise, 0.1% Gaussian noise and 20% Laplacien noise into the watermarked object as shown in Figs. 13, 14 and 15. The watermark can still be retrieved successfully, and the responses of the watermark detector are 0.5463 , 0.1176 and 0.3080 .

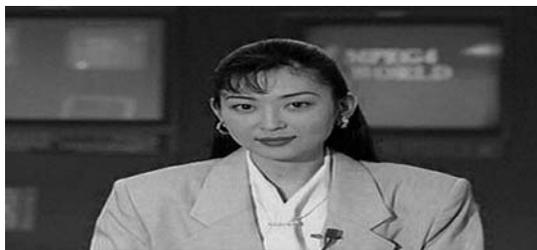

Fig. 13  Watermarked object 'Akiyo' after 30% uniform  noise adding.

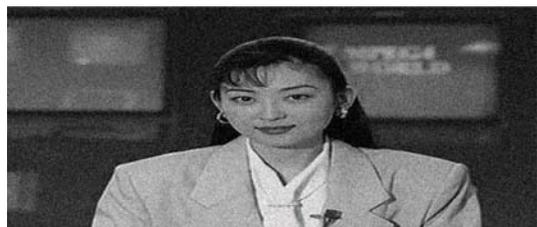

Fig. 14 Watermarked object 'Akiyo' after 0.1% Gaussian noise adding.

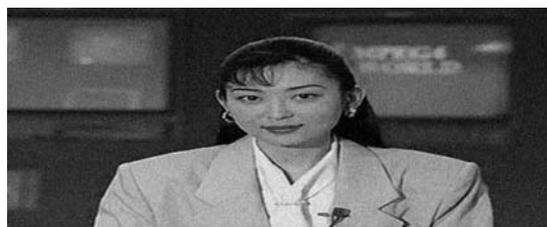

Fig. 15  Watermarked object 'Akiyo' after 20% Laplacien noise adding.

### 4.4  Filtering

Filtering is very common in image processing. The watermarked object was filtered with 3×3 blur filter and 5×5 median filter (see Fig.16 and 17). The responses of the watermark detector are 0.2248 and 0.2428. These responses are well above the threshold $Tc$ , even if the objects appeared degraded.

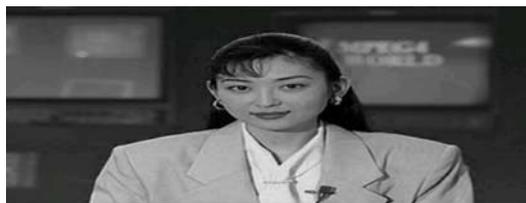

Fig. 16 Watermarked object 'Akiyo' after 3 × 3 blur filtering

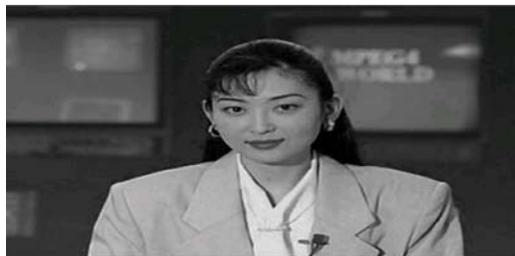

Fig. 17  Watermarked object 'Akiyo' after 5 × 5 median filtering

### 4.5  Rescaling

Scaling is also very easy to perform during the editing of digital images. So the watermark technique must be robust to the scaling attack. We test our scheme in the case of scaling the watermarked object by 0.5× 0.5. The experiment results show the watermark can still be retrieved as shown in Fig.18 et 19 with the detector response 0.1308.

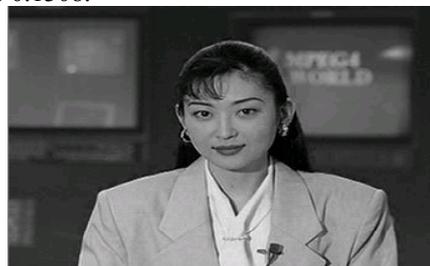

Fig. 18  Watermarked object 'Akiyo' after Rescaling   0.5 × 0.5 rescaling (50%)

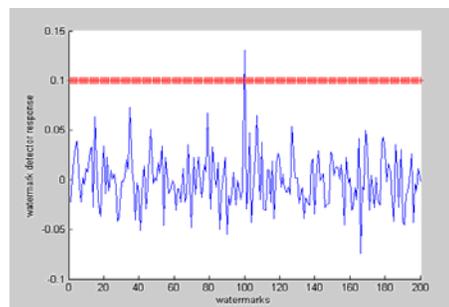

Fig. 19  Detector response of  0.5 × 0.5 rescaling

### 4.6  Multiple Watermarking Attack

The original object is watermarked, then the watermarked object is again watermarked, and so on until the object with different watermarks is obtained (see Fig. 20). The detector performs well in retrieving all the two watermarks embedded in the object, as shown in Fig. 21.

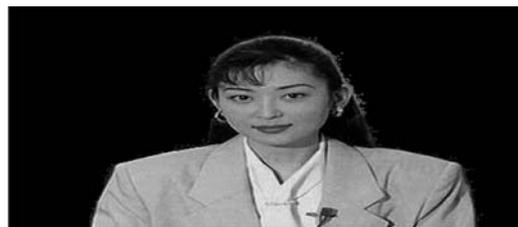

Fig. 20 The object 'Akiyo' with two different    watermarks.

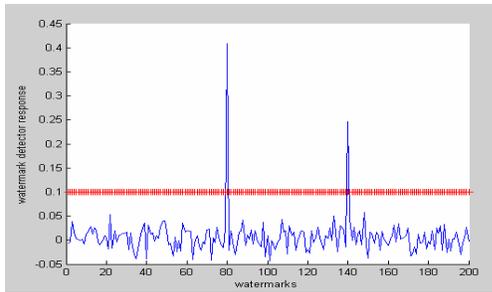

Fig. 21 Detector response of the multiple watermarked object 'Akiyo'

Table 2: Watermark detector responses after attacks

| Detector responses | akiyo | News | Sean |
|---|---|---|---|
| Sans attack | 0.7101 | 0.7212 | 0.7131 |
| JPEG quality 65% | 0.2435 | 0.3535 | 0.3944 |
| JPEG quality 70% | 0.3102 | 0.4077 | 0.4606 |
| JPEG quality 85% | 0.4676 | 0.5253 | 0.5962 |
| JPEG2000 quality 65% | 0.3491 | 0.3652 | 0.5245 |
| JPEG2000 quality 75% | 0.4231 | 0.4179 | 0.6481 |
| JPEG2000 quality 85% | 0.5296 | 0.5499 | 0.6922 |
| Uniform noise 10% | 0.6673 | 0.6798 | 0.7022 |
| Uniform noise 20% | 0.6024 | 0.6266 | 0.6470 |
| Uniform noise 30% | 0.5338 | 0.5693 | 0.6040 |
| Uniform noise 100% | 0.1556 | 0.2510 | 0.2984 |
| Laplacien noise 20% | 0.2298 | 0.5478 | 0.3867 |
| Gaussian noise 10% | 0.1357 | 0.1018 | 0.1307 |
| Blur filtering 3 * 3 | 0.1983 | 0.1714 | 0.2752 |
| Median filtering 5 * 5 | 0.2224 | 0.3790 | 0.4175 |
| Gaussian filtering | 0.6404 | 0.6726 | 0.6801 |
| Scaling 75% | 0.2382 | 0.2072 | 0.2918 |
| Scaling 50% | 0.1084 | 0.1161 | 0.2190 |

## 5. Conclusions

This paper proposes a watermarking scheme that can embed a watermark to an arbitrarily shaped object in an image. Before embedding, the image owner specifies an object of arbitrary shape that is of a concern to him. Then the object is transformed into the wavelet domain using the in place shape adaptive wavelet transform and a watermark is embedded by modifying the wavelet coefficients. Experimental results show that this scheme is robust to common signal processing procedures such as compression, median filtering and additive noise.

Efficiency of the method is revealed on the basis of the following results:

(1) The average has a smaller change than that of individual coefficient. Thus, unlike most watermarking schemes, the watermark is not embedded by just an individual wavelet coefficient but by modulating the average of the wavelet blocks.

(2) Visual model allowed to achieve the best tradeoff between transparency and robustness.

(3) Watermark detection is accomplished without the original.

(4) Many parameters can be used as private key to that they are unknown to public.

(5) This algorithm can be used for MPEG4 video watermarking

The proposed scheme was compared to the Xiangwei Kong's scheme [6]. The insufficiencies of our proposed approach are that robustness is relatively inferior to the Gaussian noise that needs exploration in our work in future.